\documentclass[a4paper]{article}

\usepackage{stackengine}
\stackMath

\usepackage{rotating}
\usepackage{bussproofs}
\usepackage{pdflscape}
\usepackage{mathabx}
\usepackage{latexsym}
\usepackage{url}
\usepackage{natbib}
\usepackage{pxfonts}
\usepackage{microtype}

\newtheorem{theorem}{Theorem}

\begin{document}

\title{A Binary Quantifier for Definite Descriptions in Intuitionist Negative Free Logic: Natural Deduction and Normalisation}
\author{Nils K\"urbis}
\date{}
\maketitle

\begin{center}
Published in the \emph{Bulletin of the Section of Logic} 48/2 (2019): 81-97
\url{http://dx.doi.org/10.18778/0138-0680.48.2.01}\bigskip
\end{center}

\begin{abstract}
\noindent This paper presents a way of formalising definite descriptions with a binary quantifier $\iota$, where $\iota x[F, G]$ is read as `The $F$ is $G$'. Introduction and elimination rules for $\iota$ in a system of intuitionist negative free logic are formulated. Procedures for removing maximal formulas of the form $\iota x[F, G]$ are given, and it is shown that deductions in the system can be brought into normal form. 
\end{abstract}

\section{Introduction}
The definite description operator $\iota$, the formal analogue of the definite article `the', is usually taken to be a term forming operator: if $A$ is a predicate, then $\iota xA$ is a term denoting the sole $A$, if there is one, or nothing or an arbitrary object if there is no or more than one $A$. This paper follows a different approach to definite descriptions by formalising them instead with a primitive binary quantifier: $\iota$ forms a formula from two predicates, and `The F is G' is formalised as $\iota x[Fx, Gx]$. The notation, and the way of treating definite descriptions that comes with it, was suggested by Dummett \citep[p.162]{dummettfregelanguage}.\footnote{Bostock considers a similar approach and explains definite descriptions as a special case of restricted quantification, where the restriction is to a single object. \citep[Sec. 8.4]{bostockintermediate} Bostock writes $(Ix\!: Fx)\ Gx$ for `The $F$ is $G$', but prefers to treat definite descriptions with a term forming operator. I owe the reference to Bostock to a referee for this journal, who also pointed me to the paper by Scott to be referred to in footnote \ref{scott} and made valuable comments on this paper.} 

The current paper treats definite descriptions purely proof theoretically. The proof theory of a term forming $\iota$ operator has been investigated in the context of sequent calculi for classical free logic by Indrzejczak \cite{andrzejmodaldescription, andrzejfregean}. Tennant gives rules for such an operator in natural deduction \citep[p.110]{tennantabstraction}.\footnote{Tennant is not explicit whether the logic in this paper is classical or intuitionist. However, as he is partial to anti-realism and constructive mathematics, we are justified in assuming that his preferred route is to add these rules to a system of intuitionist free logic. The rules are also in \citep[Ch. 7.10]{tennantnatural}, where the logic is classical.} The approach followed here may be new to the literature. 

In this paper, I investigate the binary quantifier $\iota$ in the context of a system of natural deduction for  intuitionist negative free logic. The application of the present treatment of definite descriptions to other systems of logic and their comparisons to systems known from the literature are left for further papers. To anticipate, using a negative free logic, the approach proposed here lends itself to a natural formalisation of a Russellian theory of definite descriptions, while it provides a natural formalisation of Lambert's minimal theory of definite descriptions when the logic is positive and free. 

First, notation. I will use $A_t^x$ to denote the result of replacing all free occurrences of the variable $x$ in the formula $A$ by the term $t$ or the result of substituting $t$ for the free variable $x$ in $A$. $t$ is free for $x$ in $A$ means that no (free) occurrences of a variable in $t$ become bound by a quantifier in $A$ after substitution. In using the notation $A_t^x$ I assume that $t$ is free for $x$ in $A$ or that the bound variables of $A$ have been renamed to allow for substitution without `clashes' of variables, but for clarity I also often mention the condition that $t$ is free for $x$ in $A$ explicitly. I also use the notation $Ax$ to indicate that $x$ is free in $A$, and $At$ for the result of substituting $t$ for $x$ in $A$.

\section{Natural Deduction for $\iota$ in Intuitionist Logic}
The introduction and elimination rules for the propositional logical constants of intuitionist logic \textbf{I} are: 

\begin{center}
\AxiomC{$A$} 
\AxiomC{$B$}
\LeftLabel{$\land  I$: \ }
\BinaryInfC{$A\land  B$}
\DisplayProof\qquad\qquad 
\AxiomC{$A\land  B$}
\LeftLabel{$\land  E$: \ }
\UnaryInfC{$A$}
\DisplayProof\qquad 
\AxiomC{$A\land  B$} 
\UnaryInfC{$B$}
\DisplayProof
\end{center}

\begin{center}
\RightLabel{$_i$}
\AxiomC{ }
\UnaryInfC{$A$}
\noLine
\UnaryInfC{$\Pi$}
\noLine
\UnaryInfC{$B$}
\RightLabel{$_i$}
\LeftLabel{$\rightarrow I$: \ }
\UnaryInfC{$A\rightarrow B$}
\DisplayProof \qquad \qquad
\AxiomC{$A\rightarrow B$}
\AxiomC{$A$}
\LeftLabel{$\rightarrow E$: \ }
\BinaryInfC{$B$}
\DisplayProof
\end{center}

\begin{center}
\AxiomC{$A$}
\LeftLabel{$\lor I$: \ }
\UnaryInfC{$A\lor B$}
\DisplayProof\qquad
\AxiomC{$B$}
\UnaryInfC{$A\lor B$}
\DisplayProof\qquad\qquad
\AxiomC{$A\lor B$}
\AxiomC{}
\RightLabel{$_i$}
\UnaryInfC{$A$}
\noLine
\UnaryInfC{$\Pi$}
\noLine
\UnaryInfC{$C$}
\AxiomC{}
\RightLabel{$_i$}
\UnaryInfC{$B$}
\noLine
\UnaryInfC{$\Sigma$}
\noLine
\UnaryInfC{$C$}
\RightLabel{$_i$}
\LeftLabel{$\lor E$: \ }
\TrinaryInfC{$C$}
\DisplayProof
\end{center}

\begin{prooftree}
\AxiomC{$\bot$}
\LeftLabel{$\bot E$: \ }
\UnaryInfC{$B$}
\end{prooftree}

\noindent where the conclusion $B$ of $\bot E$ is restricted to atomic formulas. 

The introduction and elimination rules for the quantifiers of \textbf{I} are: 

\begin{center}
\AxiomC{$A_y^x$}
\LeftLabel{$\forall I$: \ }
\UnaryInfC{$\forall x A$}
\DisplayProof\qquad\qquad
\AxiomC{$\forall xA$}
\LeftLabel{$\forall E$: \ }
\UnaryInfC{$A_t^x$}
\DisplayProof
\end{center} 

\noindent where in $\forall I$, $y$ is not free in any undischarged assumptions that $A_y^x$ depends on, and either $y$ is the same as $x$ or $y$ is not free in $A$; and in $\forall E$, $t$ is free for $x$ in $A$. 

\begin{center}
\AxiomC{$A_t^x$}
\LeftLabel{$\exists I$: \ }
\UnaryInfC{$\exists x A$}
\DisplayProof\qquad\qquad
\AxiomC{$\exists xA$}
\AxiomC{}
\RightLabel{$_i$}
\UnaryInfC{$A_y^x$}
\noLine
\UnaryInfC{$\Pi$}
\noLine
\UnaryInfC{$C$}
\RightLabel{$_i$}
\LeftLabel{$\exists E$: \ }
\BinaryInfC{$C$}
\DisplayProof
\end{center} 

\noindent where in $\exists I$, $t$ is free for $x$ in $A$; and in $\exists E$, $y$ is not free in $C$ nor any undischarged assumptions it depends on in $\Pi$ except $A_y^x$, and either $y$ is the same as $x$ or it is not free in $A$. 

The introduction and elimination rules for identity are: 

\begin{center}
\AxiomC{}
\LeftLabel{$=I$: \ }
\UnaryInfC{$t=t$}
\DisplayProof\qquad\qquad
\AxiomC{$t_1=t_2$}
\AxiomC{$A_{t_1}^x$}
\LeftLabel{$= E$: \ } 
\BinaryInfC{$A_{t_2}^x$}
\DisplayProof
\end{center} 

\noindent where $A$ is an atomic formula. To exclude vacuous applications of $=E$, we can require that $x$ is free in $A$ and that $t_1$ and $t_2$ are different. An induction over the complexity of formulas shows that the rule holds for formulas of any complexity. 

To formalise definite descriptions, one could add the binary quantifier $\iota$ to \textbf{I}. Its introduction and elimination rules would be: 

\begin{prooftree}
\AxiomC{$F^x_t$}
\AxiomC{$G^x_t$}
\AxiomC{}
\RightLabel{$_i$}
\UnaryInfC{$F^x_z$}
\noLine
\UnaryInfC{$\Pi$}
\noLine
\UnaryInfC{$z=t$}
\RightLabel{$_i$}
\LeftLabel{$\iota I:$ \ }
\TrinaryInfC{$\iota x[F, G]$}
\end{prooftree}

\noindent where $t$ is free for $x$ in $F$ and in $G$, and $z$ is different from $x$, not free in $t$ and does not occur free in any undischarged assumptions in $\Pi$ except $F_z^x$.\footnote{A more precise and general statement of the introduction rule for $\iota$ would result if we were to require $\Pi$ to be a deduction of $(y=t)_z^y$ from $(F_y^x)_z^y$, where $y$ is different from $x$ and not free in $t$, and either $z$ is the same as $y$ or $z$ is not free in $F_y^x$ nor in $y=t$.} 

\begin{prooftree}
\AxiomC{$\iota x[F, G]$}
\AxiomC{$\underbrace{\overline{ \ F_z^x\ }^i, \ \overline{ \ G_z^x \ }^i}$}
\noLine
\UnaryInfC{$\Pi$}
\noLine
\UnaryInfC{$C$}
\LeftLabel{$\iota E^1:$ \ }
\RightLabel{$_i$}
\BinaryInfC{$C$}
\end{prooftree}

\noindent where $z$ is not free in $C$ nor any undischarged assumptions it depends on except $F_z^x$ and $G_z^x$, and either $z$ is the same as $x$ or it is not free in $F$ nor in $G$.

\begin{prooftree}
\AxiomC{$\iota x[F, G]$}
\AxiomC{$F_{t_1}^x$}
\AxiomC{$F_{t_2}^x$}
\LeftLabel{$\iota E^2:$ \ }
\TrinaryInfC{$t_1=t_2$}
\end{prooftree}

\noindent where $t_1$ and $t_2$ are free for $x$ in $F$.\bigskip 

\noindent For simplicity we could require that $x$ occurs free in $F$ and $G$. If we don't, the truth or falsity of $\iota x[F, G]$ may depend on properties of the domain of quantification: if $F$ is true and does not contain $x$ free, then $\iota x[F, G]$ is false if there is more than one thing in the domain of quantification, and it is true if there is only one thing and $G$ is true (of the one thing, if $x$ is free in $G$). 

$\iota x[F, G]$ and $\exists x (F\land\forall y(F_y^x\rightarrow y=x)\land G)$ are interderivable. Notice that the rules for identity are not applied in the two deductions to follow.\bigskip

\noindent 1. $\iota x[F, G]\vdash \exists x (F\land\forall y(F_y^x\rightarrow y=x)\land G)$\bigskip

\noindent Let $y$ be different from $x$ and not free in $F$ or $G$:  

\begin{prooftree}
\AxiomC{$\iota x[F, G]$}
\AxiomC{$\iota x[F, G]$}
\AxiomC{}
\RightLabel{$_1$}
\UnaryInfC{$F_y^x$}
\AxiomC{}
\RightLabel{$_2$}
\UnaryInfC{$F$}
\RightLabel{$_{\iota E^2}$}
\TrinaryInfC{$y=x$}
\RightLabel{$_1$}
\UnaryInfC{$(F_y^x\rightarrow y=x)$}
\UnaryInfC{$\forall y(F_y^x\rightarrow y=x)$}
\AxiomC{}
\RightLabel{$_2$}
\UnaryInfC{$F$}
\BinaryInfC{$F\land \forall y(F_y^x\rightarrow y=x)$}
\AxiomC{}
\RightLabel{$_2$}
\UnaryInfC{$G$}
\BinaryInfC{$(F\land\forall y(F_y^x\rightarrow y=x)\land G)$}
\UnaryInfC{$\exists x (F\land\forall y(F_y^x\rightarrow y=x)\land G)$}
\RightLabel{$_{2 \ \iota E^1}$}
\BinaryInfC{$\exists x (F\land\forall y(F_y^x\rightarrow y=x)\land G)$}
\end{prooftree}

\noindent 2. $\exists x (F\land\forall y(F_y^x\rightarrow y=x)\land G)\vdash\iota x[F, G]$\bigskip

\noindent Let $y$ be different from $x$ and not free in $F$ or $G$, and let $\bigocoasterisk$ be the formula $(F\land\forall y(F_y^x\rightarrow y=x)\land G)$: 

\begin{prooftree}
\AxiomC{$\exists x\bigocoasterisk$}
\AxiomC{}
\RightLabel{$_2$}
\UnaryInfC{$\bigocoasterisk$}
\UnaryInfC{$F$}
\AxiomC{}
\RightLabel{$_2$}
\UnaryInfC{$\bigocoasterisk$}
\UnaryInfC{$G$}
\AxiomC{}
\RightLabel{$_2$}
\UnaryInfC{$\bigocoasterisk$}
\UnaryInfC{$\forall y(F_y^x\rightarrow y=x)$}
\UnaryInfC{$F_y^x\rightarrow y=x$}
\AxiomC{}
\RightLabel{$_1$}
\UnaryInfC{$F_y^x$}
\BinaryInfC{$y=x$}
\RightLabel{$_{1 \ \iota I}$}
\TrinaryInfC{$\iota x[F, G]$}
\RightLabel{$_2$}
\BinaryInfC{$\iota x[F, G]$}
\end{prooftree}

\section{Intuitionist Free Logic}
It is more interesting to add the $\iota$ quantifier to a free logic. I will use formalisations of intuitionist free logic with a primitive predicate $\exists !$, to be interpreted as `$x$ exists' or `$x$ refers' or `$x$ denotes'.  The introduction and elimination rules for the quantifiers are:\bigskip 

\begin{center}
\AxiomC{}
\RightLabel{$_i$}
\UnaryInfC{$\exists !y$}
\noLine
\UnaryInfC{$\Pi$}
\noLine
\UnaryInfC{$A_y^x$}
\LeftLabel{$\forall I:$ \ }
\RightLabel{$_i$}
\UnaryInfC{$\forall x A$}
\DisplayProof\qquad\qquad
\AxiomC{$\forall xA$}
\AxiomC{$\exists !t$}
\LeftLabel{$\forall E:$ \ }
\BinaryInfC{$A_t^x$}
\DisplayProof
\end{center} 

\noindent where in $\forall I$, $y$ does not occur free in any undischarged assumptions of $\Pi$ except $\exists !y$, and either $y$ is the same as $x$ or $y$ is not free in $A$; and in $\forall E$, $t$ is free for $x$ in $A$. 

\begin{center}
\AxiomC{$A_t^x$}
\AxiomC{$\exists !t$}
\LeftLabel{$\exists I:$ \ }
\BinaryInfC{$\exists x A$}
\DisplayProof\qquad\qquad
\AxiomC{$\exists xA$}
\AxiomC{$\underbrace{\overline{ \ A_y^x \ }^i, \overline{ \ \exists !y \ }^i}$}
\noLine
\UnaryInfC{$\Pi$}
\noLine
\UnaryInfC{$C$}
\RightLabel{$_i$}
\LeftLabel{$\exists E:$ \ }
\BinaryInfC{$C$}
\DisplayProof
\end{center} 

\noindent where in $\exists I$, $t$ is free for $x$ in $A$; and in $\exists E$, $y$ is not free in $C$ nor any undischarged assumptions of $\Pi$, except $A_y^x$ and $\exists ! y$, and either $y$ is the same as $x$ or it is not free in $A$. 

The elimination rule for identity in intuitionist free logic is the same as in \textbf{I}. 

In \emph{intuitionist positive free logic} \textbf{IPF}, identity has the same introduction rule as in intuitionist logic, i.e. $\vdash t=t$, for any term $t$. Semantically speaking, in positive free logic any statement of self-identity is true, irrespective of whether a term refers or not. 

In \emph{intuitionist negative free logic} \textbf{INF} the introduction rule for identity is weakened and requires an existential premise: 

\begin{prooftree}
\AxiomC{$\exists ! t$}
\LeftLabel{$= I^n:$ \ }
\UnaryInfC{$t=t$}
\end{prooftree}

\noindent In \textbf{INF} the existence of $t_i$ may be inferred if $t_i$ occurs in an atomic formula: 

\begin{prooftree}
\AxiomC{$At_1\ldots t_n$}
\LeftLabel{$AD:$ \ }
\UnaryInfC{$\exists ! t_i$}
\end{prooftree}

\noindent where $A$ is an $n$-place predicate letter (including identity) and $1\leq i\leq n$. Speaking semantically, for an atomic sentence, including identities, to be true, all terms in it must refer. If the language has function symbols, there is also the rule of functional denotation:  

\begin{prooftree}
\AxiomC{$\exists ! ft_1\ldots t_n$}
\LeftLabel{$FD:$ \ }
\UnaryInfC{$\exists ! t_i$}
\end{prooftree}

\noindent where $f$ is an $n$-place function letter and $1\leq i\leq n$. Speaking semantically, for the value of a function to exist, all of its arguments must exist. $=I^n$, $AD$ and $FD$ are called the rules of strictness.\footnote{\label{scott}\textbf{INF} is the system introduced by Scott \citep{scottidentityexistence} and called \textbf{Nie} by Troelstra and Schwichtenberg \citep[200]{troelstraschwichtenberg}, but with a simpler theory of identity. It is the system that results if classical \emph{reductio ad absurdum}, the rule that licenses the derivation of $A$ if $\neg A$ entails a contradiction, is not taken to form part of the system Tennant presents in \citep[Ch. 7.10]{tennantnatural}.}

Hintikka's Law $\exists !t\leftrightarrow\exists x \ x=t$, where $x$ not in $t$, is provable in \textbf{INF} and \textbf{IPF}. In \textbf{IPF}, it suffices to observe the following: 

\begin{prooftree}
\AxiomC{}
\UnaryInfC{$t=t$}
\AxiomC{$\exists ! t$}
\BinaryInfC{$\exists x \ x=t$}
\DisplayProof\qquad\qquad
\AxiomC{$\exists x \ x=t$}
\AxiomC{}
\RightLabel{$_1$}
\UnaryInfC{$x=t$}
\AxiomC{}
\RightLabel{$_1$}
\UnaryInfC{$\exists !x$}
\BinaryInfC{$\exists ! t$}
\RightLabel{$_1$}
\BinaryInfC{$\exists !t$}
\end{prooftree}

\noindent In \textbf{INF}, conclude $t=t$ from $\exists ! t$. 

The degree of a formula is the number of connectives occurring in it. $\bot$, being a connective, is of degree 1. This excludes the superfluous case in which $\bot$ is inferred from $\bot$ by $\bot E$. $\exists !t$ is an atomic formula of degree $0$. 

The \emph{major premise} of an elimination rule is the premise with the connective that the rule governs. The other premises are minor premises. A \emph{maximal formula} is one that is the conclusion of an introduction rule and the major premise of an elimination rule for its main connective. A \emph{segment} is a sequence of formulas of the same shape, all minor premises and conclusions of $\lor E$ or $\exists E$, except the first and the last one; the first is only a minor premise, the last only a conclusion. A segment is maximal if its first formula has been derived by an application of an introduction rule for its main connective, and its last formula is the major premise of an elimination rule. A deduction is \emph{in normal form} if it contains neither maximal formulas nor maximal segments. A normalisation theorem establishes that any deduction can be brought into normal form by applying \emph{reduction procedures} for the removal of maximal formulas from deductions and \emph{permutative reduction procedures} for reducing maximal segments to maximal formulas. 

Notice that the conditions imposed on applications of $=E$ have the consequence that there are no maximal formulas of the form $t_1=t_2$. 

$AD$ and $FD$ have the characteristics of introduction rules for $\exists !$, and $=I^m$ has the characteristics of an elimination rule for it. In a sense $\forall E$ and $\exists I$ of free logic also eliminate formulas of the form $\exists !t$. I will, however, not count these rules as introduction and elimination rules for $\exists !$, as there is no general way of removing formulas of the form $\exists ! t$ that have been concluded by $AD$ or $FD$ and are premises of $=I^n$, $\forall E$ or $\exists I$. 

Proofs of the normalisation theorem for intuitionist logic, such as those given by Prawitz \citep[Ch. IV.1]{prawitznaturaldeduction} and Troelstra and Schwichtenberg \citep[Ch. 6.1]{troelstraschwichtenberg}, can be modified to carry over to the intuitionist free logics considered here. 

A normalisation theorem for intuitionist negative free logic with a term forming $\iota$ operator can be reconstructed from material Tennant provides in \cite{tennantnatural}. In particular, as in the case of \textbf{I}, we can assume that every application of $\forall I$ and $\exists E$ has its own variable, that is, the free variable $y$ of an application of such a rule occurs only in the hypotheses discharged by the rule and formulas concluded from them and, for $\forall I$, in the premise of that rule and the formulas it has been derived from. This way we avoid `clashes' between the restrictions on the variables of different application of these rules when reduction procedures are applied to a deduction containing maximal formulas. Applying the reduction procedures for quantifiers of free logic can only introduce maximal formulas of lower degree than the one removed. I leave the details to the reader.

\section{Natural Deduction for $\iota$ in INF}
The interderivability of $\iota x[F, G]$ and $\exists x (F\land\forall y(F_y^x\rightarrow x=y)\land G)$ is the hall mark of a Russellian theory of definite descriptions, in which any statement of the form `The $F$ is $G$' is false if there is no $F$ or if there is more than one. It is the generally accepted treatment of definite descriptions in negative free logic. To establish how to modify the rules for $\iota$ given in Section 2 to yield a Russellian theory of definite descriptions when the logic is  intuitionist negative free logic, we analyse the deductions establishing the interderivability of $\iota x[F, G]$ and $\exists x (F\land\forall y(F_y^x\rightarrow x=y)\land G)$ in \textbf{I} given at the end of that section. 

Looking at the derivation of $\exists x (F\land\forall y(F_y^x\rightarrow x=y)\land G)$  from $\iota x[F, G]$, had the application of the universal quantifier introduction rule be one of free logic, it would have allowed the discharge of an assumption $\exists !y$, and had the existential quantifier introduction rule been one of free logic, a further assumption $\exists !x$ would have been required. Both lend themselves as additional premises of $\iota E^2$, as premises analogous to the existence assumptions in the rules of the quantifiers of free logic. $\exists ! y$ would be discharged by the application of the universal quantifier introduction rule of free logic, so in order for the conclusion of the deduction not to depend on $\exists ! x$, it would have to be discharged, and the only option here is that it is discharged by the application of $\iota E^1$. This is also a natural option, corresponding, as it does, to the discharge of existence assumptions by the quantifier rules of free logic. 

Generalising the first observation, we add the premises $\exists !t_1$ and $\exists !t_2$ to $\iota E^2$: 

\begin{prooftree}
\AxiomC{$\iota x[F, G]$}
\AxiomC{$\exists !t_1$}
\AxiomC{$\exists ! t_2$}
\AxiomC{$F_{t_1}^x$}
\AxiomC{$F_{t_2}^x$}
\LeftLabel{$\iota E^2: \quad$}
\QuinaryInfC{$t_1=t_2$}
\end{prooftree}

\noindent where $t_1$ and $t_2$ are free for $x$ in $F$.

To implement the second observation, we add $\exists !z$ as an additional discharged assumption to $\iota E^1$: 

\begin{prooftree}
\AxiomC{$\iota x[F, G]$}
\AxiomC{$\underbrace{\overline{ \ F_z^x\ }^i, \ \overline{ \ G_z^x \ }^i, \ \overline{ \ \exists !z \ }^i}$}
\noLine
\UnaryInfC{$\Pi$}
\noLine
\UnaryInfC{$C$}
\LeftLabel{$\iota E^1: \quad$}
\RightLabel{$_i$}
\BinaryInfC{$C$}
\end{prooftree}

\noindent where is $z$ not free in $C$ nor any undischarged assumptions it depends on except $F_z^x$, $G_z^x$ and $\exists ! z$, and either $z$ is the same as $x$ or it is not free in $F$ nor in $G$.

To find suitable modifications of the introduction rule for $\iota$, we look at the derivation of $\iota x[F, G]$ from $\exists x (F\land\forall y(F_y^x\rightarrow x=y)\land G)$ in \textbf{I}. Had the application of the universal quantifier elimination rule been one of free logic, a further assumption $\exists !y$ would have been required, and had the existential quantifier elimination rule been one of free logic, it would have allowed the discharge of an assumption $\exists !x$. The latter lends itself as an additional premise of $\iota I$, the former as an additional assumption discharged by that rule, which is again analogous to the existence assumptions required and discharged in applications of the rules for the quantifiers of free logic. 

Generalising the second observation, we add $\exists !t$ as a further premise, and to implement the first observation we add $\exists !z$ as a further discharged assumption to $\iota I$: 

\begin{prooftree}
\AxiomC{$F^x_t$}
\AxiomC{$G^x_t$}
\AxiomC{$\exists !t$}
\AxiomC{$\underbrace{\overline{ \ F_z^x\ }^i, \ \overline{ \ \exists !z \ }^i}$}
\noLine
\UnaryInfC{$\Pi$}
\noLine
\UnaryInfC{$z=t$}
\RightLabel{$_i$}
\LeftLabel{$\iota I: \qquad$}
\QuaternaryInfC{$\iota x[F, G]$}
\end{prooftree}

\noindent where $t$ is free for $x$ in $F$ and in $G$, and $z$ is different from $x$, not free in $t$ and does not occur free in any undischarged assumptions in $\Pi$ except $F_z^x$ and $\exists ! z$.\footnote{A more precise and general statement of the introduction rule for $\iota$ would result if we were to require $\Pi$ to be a deduction of $(y=t)_z^y$ from $(F_y^x)_z^y$ and $\exists !z$, where $y$ is different from $x$ and not free in $t$, and either $z$ is the same as $y$ or $z$ is not free in $F_y^x$ nor in $y=t$.}

It is obvious that $\iota x[F, G]$ and $\exists x (F\land\forall y(F_y^x\rightarrow x=y)\land G)$ are interderivable in \textbf{INF} when $\iota$ is governed by the modified rules, but we give the deductions for convenience. 

\begin{landscape}
\noindent 1. $\iota x[F, G]\vdash \exists x (F\land\forall y(F_y^x\rightarrow y=x)\land G)$\bigskip

\noindent Let $x$ and $y$ be different variables, where $y$ is not free in $F$ nor in $G$: \bigskip

\begin{prooftree}
\AxiomC{$\iota x[F,G]$}
\AxiomC{$\iota x[F, G]$}
\AxiomC{}
\RightLabel{$_2$}
\UnaryInfC{$\exists ! y$}
\AxiomC{}
\RightLabel{$_3$}
\UnaryInfC{$\exists ! x$}
\AxiomC{}
\RightLabel{$_1$}
\UnaryInfC{$F_y^x$}
\AxiomC{}
\RightLabel{$_3$}
\UnaryInfC{$F$}
\RightLabel{$_{\iota E^2}$}
\QuinaryInfC{$y=x$}
\RightLabel{$_1$}
\UnaryInfC{$(F_y^x\rightarrow y=x)$}
\RightLabel{$_2$}
\UnaryInfC{$\forall y(F_y^x\rightarrow y=x)$}
\AxiomC{}
\RightLabel{$_3$}
\UnaryInfC{$F$}
\BinaryInfC{$F\land \forall y(F_y^x\rightarrow y=x)$}
\AxiomC{}
\RightLabel{$_3$}
\UnaryInfC{$G$}
\BinaryInfC{$F\land\forall y(F_y^x\rightarrow y=x)\land G$}
\AxiomC{}
\RightLabel{$_3$}
\UnaryInfC{$\exists ! x$}
\BinaryInfC{$\exists x (F\land\forall y(F_y^x\rightarrow y=x)\land G)$}
\RightLabel{$_{3 \ \iota E^1}$}
\BinaryInfC{$\exists x (F\land\forall y(F_y^x\rightarrow y=x)\land G)$}
\end{prooftree}

\noindent 2. $\exists x (F\land\forall y(F_y^x\rightarrow y=x)\land G)\vdash\iota x[F, G]$\bigskip

\noindent Let $\bigocoasterisk$ be the formula $(F\land\forall y(F_y^x\rightarrow y=x)\land G)$, where $y$ is different from $x$ and not free in $F$ or $G$: 

\begin{prooftree}
\AxiomC{$\exists x\bigocoasterisk$}
\AxiomC{}
\RightLabel{$_2$}
\UnaryInfC{$\bigocoasterisk$}
\UnaryInfC{$F$}
\AxiomC{}
\RightLabel{$_2$}
\UnaryInfC{$\bigocoasterisk$}
\UnaryInfC{$G$}
\AxiomC{}
\RightLabel{$_2$}
\UnaryInfC{$\exists ! x$}
\AxiomC{}
\RightLabel{$_2$}
\UnaryInfC{$\bigocoasterisk$}
\UnaryInfC{$\forall y(F_y^x\rightarrow y=x)$}
\AxiomC{}
\RightLabel{$_1$}
\UnaryInfC{$\exists ! y$}
\BinaryInfC{$F_y^x\rightarrow y=x$}
\AxiomC{}
\RightLabel{$_1$}
\UnaryInfC{$F_y^x$}
\BinaryInfC{$y=x$}
\RightLabel{$_{1 \ \iota I}$}
\QuaternaryInfC{$\iota x[F, G]$}
\RightLabel{$_2$}
\BinaryInfC{$\iota x[F, G]$}
\end{prooftree}

\noindent Let $\mathbf{INF}^\iota$ denote the systems of intuitionist negative free logic augmented with the rules for $\iota$ given in this section.  

In order to prove a normalisation theorem for $\mathbf{INF}^\iota$, we first observe that $\bot E$ can be restricted to atomic conclusions in this system: 

\begin{enumerate}
\item Instead of inferring $\forall xA$ from $\bot$, infer $A_y^x$, for some $y$ not occurring in any assumption that $\bot$ depends on, and apply $\forall I$, discharging vacuously. 

\item Instead of inferring $\exists xA$ from $\bot$, infer $A_t^x$, for some $t$ that is free for $x$ in $A$, infer $\exists !t$, and apply $\exists I$. 

\item Instead of inferring $\iota x[F, G]$ from $\bot$, infer $F_t^x$, $G_t^x$, $\exists ! t$ and $z=t$, for some $t$ that is free for $x$ in $F$ and in $G$ and some $z$ that is not free in any assumption that $\bot$ depends on, and apply $\iota I$, discharging vacuously. 
\end{enumerate} 

\noindent Next, $=E$ can be restricted to atomic formulas in $\mathbf{INF}^\iota$. Consider an application of this rule with premise $\iota x[F, G]_{t_1}^y$:

\begin{prooftree}
\AxiomC{$t_1=t_2$}
\AxiomC{$\iota x[F, G]_{t_1}^y$}
\BinaryInfC{$\iota x[F, G]_{t_2}^y$}
\end{prooftree}

\noindent where $t_1$ and $t_2$ are free for $y$ in $\iota x[F, G]$. The exclusion of vacuous applications of $=E$ means that $y$ must be different from $x$, and so $\iota x[F, G]_{t_1}^y$ is $\iota x[F_{t_1}^y, G_{t_1}^y]$. Let $v$ and $z$ be different variables not occurring in $F$, $G$, $t_1$, $t_2$.  The induction step applying $=E$ to subformulas of $\iota x[F, G]_{t_1}^y$ is the following: 

\begin{prooftree}
\AxiomC{$\iota x[F_{t_1}^y, G_{t_1}^y]$}
\AxiomC{$t_1=t_2$}
\AxiomC{}
\RightLabel{$_2$}
\UnaryInfC{$(F_{t_1}^y)_z^x$}
\BinaryInfC{$(F_{t_2}^y)_z^x$}
\AxiomC{$t_1=t_2$}
\AxiomC{}
\RightLabel{$_2$}
\UnaryInfC{$(G_{t_1}^y)_z^x$}
\BinaryInfC{$(G_{t_2}^y)_z^x$}
\AxiomC{}
\RightLabel{$_2$}
\UnaryInfC{$\exists ! z$}
\AxiomC{$\iota x[F_{t_1}^y, G_{t_1}^y]$}
\AxiomC{}
\RightLabel{$_1$}
\UnaryInfC{$\exists !v$}
\AxiomC{}
\RightLabel{$_2$}
\UnaryInfC{$\exists !z$}
\AxiomC{}
\RightLabel{$_1$}
\UnaryInfC{$(F_{t_1}^y)_v^x$}
\AxiomC{}
\RightLabel{$_2$}
\UnaryInfC{$(F_{t_1}^y)_z^x$}
\RightLabel{$_{\iota E^2}$}
\QuinaryInfC{$v=z$}
\RightLabel{$_{1 \ \iota I}$}
\QuaternaryInfC{$\iota x[F_{t_2}^y, G_{t_2}^y]$}
\RightLabel{$_{2 \ \iota E^1}$}
\BinaryInfC{$\iota x[F_{t_2}^y, G_{t_2}^y]$}
\end{prooftree}
\end{landscape}

As for applications of $\forall I$ and $\exists E$, we can assume that every application of $\iota I$ and $\iota E^1$ has its own free variable, i.e. the variable $z$ of an application of $\iota I$ or $\iota E^2$ occurs only in the premises discharged by the rule and formulas derived from the discharged premises, and nowhere else in the deduction. 

I will now give the reduction procedures for maximal formulas of the form $\iota x[F, G]$ and the permutative reduction procedures for maximal segments consisting of a formula of that form. 

There are two cases of reduction procedures for maximal formulas of the form $\iota x[F, G]$ to be considered. First, the conclusion of $\iota I$ is the major premise of $\iota E^1$: 

\begin{prooftree}
\AxiomC{$\Sigma_1$}
\noLine
\UnaryInfC{$F^x_t$}
\AxiomC{$\Sigma_2$}
\noLine
\UnaryInfC{$G^x_t$}
\AxiomC{$\Sigma_3$}
\noLine
\UnaryInfC{$\exists !t$}
\AxiomC{$\underbrace{\overline{ \ F_z^x\ }^i, \ \overline{ \ \exists !z \ }^i}$}
\noLine
\UnaryInfC{$\Pi$}
\noLine
\UnaryInfC{$z=t$}
\RightLabel{$_i$}
\QuaternaryInfC{$\iota x[F, G]$}
\AxiomC{$\underbrace{\overline{ \ F_v^x\ }^j, \ \overline{ \ G_v^x \ }^j, \ \overline{ \ \exists !v \ }^j}$}
\noLine
\UnaryInfC{$\Xi$}
\noLine
\UnaryInfC{$C$}
\RightLabel{$_j$}
\BinaryInfC{$C$}
\end{prooftree}

\noindent Transform such steps in a deduction into the following, where $\Xi_t^v$ is the deduction resulting from $\Xi$ by replacing the variable $v$ everywhere with the term $t$: 

\begin{prooftree}
\AxiomC{$\underbrace{\stackon{F_t^x}{\Sigma_1}, \ \stackon{G_t^x}{\Sigma_2}, \ \stackon{\exists !t}{\Sigma_3}}$}
\noLine
\UnaryInfC{$\Xi_t^v$}
\noLine
\UnaryInfC{$C$}
\end{prooftree}

\noindent The conditions on variables ensure that no clashes arise from the replacement. 

Second, the conclusion of $\iota I$ is the major premise of $\iota E^2$: 

\begin{prooftree}
\AxiomC{$\Sigma_1$}
\noLine
\UnaryInfC{$F^x_{t_1}$}
\AxiomC{$\Sigma_2$}
\noLine
\UnaryInfC{$G^x_{t_1}$}
\AxiomC{$\Sigma_3$}
\noLine
\UnaryInfC{$\exists ! t_1$}
\AxiomC{$\underbrace{\overline{ \ F_z^x\ }^i, \ \overline{ \ \exists !z \ }^i}$}
\noLine
\UnaryInfC{$\Pi$}
\noLine
\UnaryInfC{$z=t_1$}
\RightLabel{$_i$}
\QuaternaryInfC{$\iota x[F, G]$}
\AxiomC{$\Xi_1$}
\noLine
\UnaryInfC{$\exists !t_2$}
\AxiomC{$\Xi_2$}
\noLine
\UnaryInfC{$\exists ! t_3$}
\AxiomC{$\Xi_3$}
\noLine
\UnaryInfC{$F_{t_2}^x$}
\AxiomC{$\Xi_4$}
\noLine
\UnaryInfC{$F_{t_3}^x$}
\QuinaryInfC{$t_2=t_3$}
\end{prooftree}

\noindent Transform such steps in a deduction into the following, where $\Pi_{t_2}^z$ and $\Pi_{t_3}^z$ are the deductions resulting from $\Pi$ by replacing $z$ with $t_2$ and $t_3$, respectively, and the last rule is an application of $=E$: 

\begin{prooftree}
\AxiomC{$\underbrace{\stackon{F_{t_3}^x}{\Xi_4},\stackon{\exists !t_3}{\Xi_2}}$}
\noLine
\UnaryInfC{$\Pi_{t_3}^z$}
\noLine
\UnaryInfC{$t_3=t_1$}
\AxiomC{$\underbrace{\stackon{F_{t_2}^x}{\Xi_3},\stackon{\exists !t_2}{\Xi_1}}$}
\noLine
\UnaryInfC{$\Pi_{t_2}^z$}
\noLine
\UnaryInfC{$t_2=t_1$}
\BinaryInfC{$t_2=t_3$}
\end{prooftree} 

\noindent The conditions on variables ensure that no clashes arise from the replacements. 

The second reduction procedure for maximal formulas of the form $\iota x[F, G]$ is slightly unusual, as it appeals to a rule for another logical constant, i.e. identity. However, as the conclusion of $\iota E^2$ is an identity, it is to be expected that its rules may have to be appealed to in the workings of the rules for $\iota$. 

I only give two examples of permutative reduction procedures for formulas of the form $\iota x[F, G]$ that are the conclusion of $\lor E$, $\exists E$ or $\iota E^1$ and the major premise of $\iota E^1$ or $\iota E^2$. As in previous cases, clashes between variables are avoidable by choosing different variables for the applications of $\exists E$ and the elimination rules for $\iota$.

First example. The major premise of $\iota E^1$ is concluded by $\exists E$:

\begin{prooftree}
\AxiomC{$\exists vA$}
\AxiomC{}
\RightLabel{$_i$}
\UnaryInfC{$A_y^v$}
\noLine
\UnaryInfC{$\Sigma$}
\noLine
\UnaryInfC{$\iota x[F, G]$}
\RightLabel{$_i$}
\BinaryInfC{$\iota x[F, G]$}
\AxiomC{$\underbrace{\overline{ \ F_z^x\ }^j, \ \overline{ \ G_z^x \ }^j, \ \overline{ \ \exists !z \ }^j}$}
\noLine
\UnaryInfC{$\Pi$}
\noLine
\UnaryInfC{$C$}
\RightLabel{$_j$}
\BinaryInfC{$C$}
\end{prooftree}

\noindent Replace such steps in a deduction by: 

\begin{prooftree}
\AxiomC{$\exists vA$}
\AxiomC{}
\RightLabel{$_i$}
\UnaryInfC{$A_y^v$}
\noLine
\UnaryInfC{$\Sigma$}
\noLine
\UnaryInfC{$\iota x[F, G]$}
\AxiomC{$\underbrace{\overline{ \ F_z^x\ }^j, \ \overline{ \ G_z^x \ }^j, \ \overline{ \ \exists !z \ }^j}$}
\noLine
\UnaryInfC{$\Pi$}
\noLine
\UnaryInfC{$C$}
\RightLabel{$_j$}
\BinaryInfC{$C$}
\RightLabel{$_i$}
\BinaryInfC{$C$}
\end{prooftree}

\noindent Second example. The major premise of $\iota E^2$ is the conclusion of $\exists E$: 

\begin{prooftree}
\AxiomC{$\exists vA$}
\AxiomC{}
\RightLabel{$_i$}
\UnaryInfC{$A_y^v$}
\noLine
\UnaryInfC{$\Sigma$}
\noLine
\UnaryInfC{$\iota x[F, G]$}
\RightLabel{$_i$}
\BinaryInfC{$\iota x[F, G]$}
\AxiomC{$\exists !t_1$}
\AxiomC{$\exists ! t_2$}
\AxiomC{$F_{t_1}^x$}
\AxiomC{$F_{t_2}^x$}
\QuinaryInfC{$t_1=t_2$}
\end{prooftree} 

\noindent Replace such steps in a deduction by: 

\begin{prooftree}
\AxiomC{$\exists vA$}
\AxiomC{}
\RightLabel{$_i$}
\UnaryInfC{$A_y^v$}
\noLine
\UnaryInfC{$\Sigma$}
\noLine
\UnaryInfC{$\iota x[F, G]$}
\AxiomC{$\exists !t_1$}
\AxiomC{$\exists ! t_2$}
\AxiomC{$F_{t_1}^x$}
\AxiomC{$F_{t_2}^x$}
\QuinaryInfC{$t_1=t_2$}
\RightLabel{$_i$}
\BinaryInfC{$t_1=t_2$}
\end{prooftree} 

\noindent The remaining cases are similar. 

I am not counting $\iota E^2$ as an introduction rule for $=$. There is no general way of removing formulas $t_1=t_2$ concluded by $\iota E^2$ and eliminated by $=E$, as the following illustrates: 

\begin{prooftree}
\AxiomC{$\iota x[F, G]$}
\AxiomC{$\exists !t_1$}
\AxiomC{$\exists ! t_2$}
\AxiomC{$F_{t_1}^x$}
\AxiomC{$F_{t_2}^x$}
\QuinaryInfC{$t_1=t_2$}
\AxiomC{$A_{t_1}^x$}
\BinaryInfC{$A_{t_2}^x$}
\end{prooftree}

\noindent Thus there are no further maximal formulas to be considered in $\mathbf{INF}^\iota$. After the theorem, I will give an alternative second elimination rule for $\iota$ that avoids this problem. 

We have the following:  

\begin{theorem}
For any deduction $\Pi$ of $A$ from $\Gamma$ in $\mathbf{INF}^\iota$ there is a deduction of the same conclusion from some of the formulas in $\Gamma$ that is in normal form.
\end{theorem} 

\noindent \emph{Proof.} By induction over the rank of proofs. The length of a segment is the number of formulas it consists of and its degree the number of logical constants in that formula. Let a maximal formula be a maximal segment of length $1$. The \emph{rank} of a deduction is the pair $\langle d, l\rangle$, where $d$ is the highest degree of a maximal segment or $0$ if there is none, and $l$ is the sum of the lengths of maximal segments of highest degree. $\langle d, l\rangle < \langle d', l'\rangle$ iff either (i) $d<d'$ or (ii) $d=d'$ and $l<l'$. Applying the reduction procedures to a suitably chosen maximal segment of highest degree and longest length reduces the rank of a deduction.\bigskip

\noindent We can reformulate the second elimination rule for $\iota$ to incorporate an application of Leibniz's Law instead of concluding with an identity: 

\begin{prooftree}
\AxiomC{$\iota x[F, G]$}
\AxiomC{$\exists !t_1 \qquad \exists ! t_2$}
\AxiomC{$F_{t_1}^x \qquad F_{t_2}^x$}
\AxiomC{$A_{t_1}^x$}
\LeftLabel{$\iota E^{2A}: \quad$}
\QuaternaryInfC{$A_{t_2}^x$}
\end{prooftree}

\noindent $A$ can be restricted to atomic formulas, an induction over the complexity of formulas showing that the general version with $A$ a formula of any degree is admissible. Call the system resulting from $\mathbf{INF}^\iota$ by replacing $\iota E^2$ with $\iota E^{2A}$ ${\mathbf{INF}^\iota}'$. 

$\iota E^2$ and $\iota E^{2A}$ are interderivable in virtue of the rules for identity: 

\begin{enumerate} 
\item To derive $\iota E^{2A}$, given premises $\iota x[F, G]$, $\exists ! t_1$, $\exists !t_2$, $F_{t_1}^x$ and $F_{t_2}^x$, derive $t_1=t_2$ by $\iota E^2$ and apply $=E$ to it and the premise $A_{t_1}^x$ to derive $A_{t_2}^x$. 

\item To derive $\iota E^2$, let $A$ be $t_1=x$, so that $A_{t_1}^x$ is $t_1=t_1$: derive it from $\exists !t_1$ by $=I^n$, apply $\iota E^{2A}$ to derive $A_{t_2}^x$, i.e. $t_1=t_2$. 

\end{enumerate} 

\noindent Thus $\mathbf{INF}^\iota$ and ${\mathbf{INF}^\iota}'$ are equivalent.  

In ${\mathbf{INF}^\iota}'$, steps in a deduction that conclude $t_1=t_2$ by $\iota E^{2A}$ (with $t_1=t_1$ as $A_{t_1}^x$) and using it as the identity in Leibniz' Law are redundant: $\iota E^{2A}$ can instead be applied with the premise and conclusion of Leibniz' Law. Such identities can therefore be removed from deductions, and we are now at liberty to count them amongst the maximal formulas. 

If a maximal formula arises from introducing $\iota x[F, G]$ by $\iota I$ and eliminating it by $\iota E^{2A}$, we have the following situation: 

\begin{prooftree}
\AxiomC{$\Sigma_1$}
\noLine
\UnaryInfC{$F^x_{t_1}$}
\AxiomC{$\Sigma_2$}
\noLine
\UnaryInfC{$G^x_{t_1}$}
\AxiomC{$\Sigma_3$}
\noLine
\UnaryInfC{$\exists ! t_1$}
\AxiomC{$\underbrace{\overline{ \ F_z^x\ }^i, \ \overline{ \ \exists !z \ }^i}$}
\noLine
\UnaryInfC{$\Pi$}
\noLine
\UnaryInfC{$z=t_1$}
\RightLabel{$_i$}
\QuaternaryInfC{$\iota x[F, G]$}
\AxiomC{$\stackon{\exists !t_2}{\Xi_1} \qquad \stackon{\exists !t_3}{\Xi_2}$}
\AxiomC{$\stackon{F_{t_2}^x}{\Xi_3} \qquad \stackon{F_{t_3}^x}{\Xi_4}$}
\AxiomC{$\stackon{A_{t_2}^x}{\Xi_5}$}
\QuaternaryInfC{$A_{t_3}^x$}
\end{prooftree}

\noindent We now have two options for removing the maximal formula. We can proceed as previously: conclude $t_2=t_3$ by an application of Leibniz' Law to the conclusions $t_2=t_1$ of $\Pi_{t_1}^z$ and $t_3=t_1$ of $\Pi_{t_2}^z$, and then apply Leibniz' Law once more with $A_{t_2}^x$ as further premise and $A_{t_3}^x$ as conclusion. Alternatively, we can first conclude $A_{t_1}^x$ from the conclusion $t_2=t_1$ of $\Pi_{t_1}^z$ and $A_{t_2}^x$, and then conclude $A_{t_3}^x$ from $A_{t_2}^x$ and the conclusion $t_3=t_1$ of $\Pi_{t_2}^z$. Thus deductions in the system resulting by replacing $\iota E^2$ by $\iota E^{2A}$ also normalise, and it has the additional advantage of avoiding identities concluded by $\iota E^2$ and eliminated by Leibniz' Law. 

Thus we have the following: 

\begin{theorem}
For any deduction $\Pi$ of $A$ from $\Gamma$ in ${\mathbf{INF}^\iota}'$ there is a deduction of the same conclusion from some formulas in $\Gamma$ that is in normal form.
\end{theorem} 

\noindent Deductions in ${\mathbf{INF}^\iota} '$ have slightly neater proof-theoretic properties than those in $\mathbf{INF}^\iota$, as deductions in normal form in ${\mathbf{INF}^\iota} '$ do not contain redundant identities introduced by $\iota E^2$ and eliminated by $=E$. Deductions in $\mathbf{INF}^\iota$ are, however, slightly simpler if we are interested in establishing identities, and this will be the case if we are interested in comparing the present system with the standard treatment of $\iota$ as a term forming operator: axioms and rules for the latter invariably appeal to identity. 

\bigskip

\setlength{\bibsep}{0pt}
\bibliographystyle{chicago}
\bibliography{iota}

\end{document}